\documentclass[lettersize,journal]{IEEEtran}
\usepackage{amsmath,amsfonts}
\usepackage{algorithmic}
\usepackage{algorithm}
\usepackage{array}
\usepackage[caption=false,font=normalsize,labelfont=sf,textfont=sf]{subfig}
\usepackage{textcomp}
\usepackage{stfloats}
\usepackage{url}
\usepackage{verbatim}
\usepackage{graphicx}
\usepackage{cite}
\usepackage{color}
\begin{document}

\title{Low-noise Fourier Transform Spectroscopy Enabled by Superconducting On-Chip Filterbank Spectrometers}

\author{Chris S. Benson\IEEEauthorrefmark{1}$^\prime$, Peter S. Barry\IEEEauthorrefmark{1}, Patrick Ashworth\IEEEauthorrefmark{1}, Harry Gordon-Moys\IEEEauthorrefmark{1}, Kirit S. Karkare\IEEEauthorrefmark{2}, Izaak Morris\IEEEauthorrefmark{1}, Gethin Robson\IEEEauthorrefmark{1}
\IEEEauthorblockA{\IEEEauthorrefmark{1}School of Physics and Astronomy, Astronomy Instrumentation Group, Cardiff University, Cardiff, UK.}
\IEEEauthorblockA{\IEEEauthorrefmark{2}Department of Physics, Boston University, Boston, MA, U.S.A.}
\thanks{$^\prime$BensonC@Cardiff.ac.uk}
\thanks{Manuscript received September, 2025; Accepted April 2026. For the purpose of open access, the author has applied a Creative Commons Attribution (CC
BY) license to any Author Accepted Manuscript version arising.}}

\markboth{IEEE Transactions on Applied Superconductivity. Accepted pre-print version.}%
{This is the author's accdepted version which has not been fully edited and content may change prior to final publication.}


\maketitle
\begin{abstract}
Spectroscopic architectures that have historically been used for large field of view mapping spectroscopy in millimetere and sub-millimetre astronomy suffer from significant drawbacks. On-chip filterbank spectrometers are a promising technology in this respect; however, they must overcome an orders-of-magnitude increase in detector counts, efficiency loss due to dielectric properties, and stringent fabrication tolerances that currently limit scaling to resolutions of order 1000 over a large array. We propose coupling a medium-resolution Fourier transform spectrometer to a low-resolution filterbank spectrometer focal plane, which serves as a post-dispersion element. In this arrangement, medium resolution imaging spectroscopy is provided by the Fourier transform spectrometer, while the low resolution filterbank spectrometer serves to decrease the photon noise inherent in typical broadband Fourier transform spectrometer measurements by over an order of magnitude. This is achieved while maintaining the excellent imaging advantages of both architectures. We present predicted mapping speeds for a filterbank-dispersed Fourier transform spectrometer from a ground-based site and a balloon-borne platform. We also demonstrate the potential that an instrument of this type has for an $R\sim1000$ line intensity mapping experiment using the James Clerk Maxwell Telescope as an example platform. We demonstrate that a filterbank-dispersed Fourier transform spectrometer would be capable of $R\sim1000$ measurements of CO power spectra with a signal-to-noise ratio of 10--100 with surveys of $10^5$--$10^6$ spectrometer hours.
\end{abstract}

\begin{IEEEkeywords}
Millimeter wave spectroscopy, Submillimeter wave spectroscopy, Fourier spectroscopy, Superconducting integrated circuits, Astronomy, Millimeter wave detectors, Submillimeter wave detectors.
\end{IEEEkeywords}

\section{Introduction}
Medium resolution ($R=\lambda/\Delta\lambda=\nu/\Delta\nu\sim1000$.) mapping spectroscopy capable of coupling to the current and next generation large field of view (FoV) focal planes has been identified as a priority for the millimetre/sub-millimeter (mm) astronomy community~\cite{Farrah2019,atLASTScience,klaassen2025,2024UKRoadmap}. Galaxy surveys and line intensity mapping (LIM) are notable areas that stand to benefit from significant advances in instruments of this type. Specifically, the Atacama Large Aperture Submillimeter Telescope (AtLAST) is targeting a mapping spectrometer with $>$1000 spatial pixels capable of resolving powers $R\sim2000$ in the 2030s to identify the ``first forming galaxies'' at redshifts, $z>10-15$~\cite{Kohno2020}. Currently, LIM experiments in the mm/sub-mm experiments seek to provide competitive constraints on expansion history, neutrinos, and the dynamics of reionisation through spatially unresolved three-dimensional measurements of galaxy distributions with sensitive mapping spectrometers~\cite{Kovetz2017,Silva2021,Karkare2022,Karkare2022_2}. An ideal LIM experiment prioritises spectrometers with high mapping speeds and large fields of view (FoVs) that target a resolving power $R\sim1000$. This has led to several mm/sub-mm LIM experiments demonstrating the capabilities of different spectrometer architectures\cite{vieira2019,2020Concerto,Karkare2022}.

Historically, both grating spectrometers and Fourier transform spectrometers (FTSs) have been employed for low to medium resolution mapping spectroscopy in the mm/sub-mm. However, both spectrometer architectures suffer significant drawbacks~\cite{Farrah2019}. 

FTSs have been developed for astronomical imaging spectroscopy capable of resolving powers as high as $R\sim5000$ with high throughput~\cite{Gom2010}. Additionally, FTSs provide access to a broad free spectral range (FSR), in practice being only limited by the performance of the optics employed (often the beamsplitter). Thus, in the mm/sub-mm, FTSs covering multiple octaves of spectral bandwidth are achievable~\cite{2008Griffin,myThesis}. Also of note for wide-field imaging spectroscopy, FTSs have a particular multiplexing advantage in that every detector at the focal plane measures a full spectrum. In spite of these capabilities, the inherent multiplexing advantage of FTSs results in a significant increase in photon noise (being dependent on the spectral bandwidth of the light incident on a direct detector).

If the inherent noise of an FTS from spectral multiplexing can be reduced, it holds significant advantages over grating spectrometers, which struggle to scale to $R\geq1000$ in the mm/sub-mm. The throughput disadvantage as a consequence of the slit apertures employed by gratings at medium to high resolution is thoroughly examined by Jacquinot (1960)~\cite{Jacquinot1960}. Here we note that the throughput of a spectrometer limited only by a circular input aperture, such as an FTS, is given as
\begin{equation}
    \Theta_I = \frac{2\pi A_I}{R},
\end{equation}
where $A_I$ is the effective area of the interferometer beam and $R$ is the resolving power of the spectrometer, while for an optimal slit-fed grating spectrometer it is given as
\begin{equation}
    \Theta_G = \frac{A_G \beta (\sin i_1 + \sin i_2)}{R},
\end{equation}
where $A_G$ is the effective area of the beam in the grating spectrometer, $\beta$ is the angular height of the slit, and $i_1$ and $i_2$ are the angles of incidence and emergence, respectively. From these, we see that the throughput of an interferometric spectrometer exceeds that of such a grating spectrometer (with equal area and resolution) by over an order of magnitude for all practical physical dimensions of such a grating. It should also be noted that spectrum from a diffraction grating is dispersed along a spatial dimension of a focal plane, making it challenging to couple a grating to a large FoV~\cite{Farrah2019}.

Fabry-Perot Spectrometers (FPSs), in a similar manner to FTSs, have a throughput advantage over grating spectrometers. Additionally they are easily capable of resolving powers of order $10^4$. However, FPS are inherently narrow banded as increasing the FSR of an FPS with a given resolution requires increasing the finesse of the resonant cavity. In an ideal FPS, finesse is dictated by the reflectivity of the cavity. However, it should be noted that the acceptance of off-axis rays from a large FoV and defects in the cavity result in a diminished throughput that scales with the reflective finesse of the cavity~\cite{Baker1982}. In practice, FPSs have a relatively narrow instantaneous free spectral range, being of order 1/20th of an octave at $R=100$ and scaling with $1/R$. As a consequence, FPSs require a spectral scanning mechanism to cover an appreciable spectral range and the precision of the actuation mechanism becomes a significant restriction on an FPS~\cite{Farrah2019}. 

Here we note an important subtlety; though both FTSs and FPIs employ a scanning mechanism, in an FTS all spectral components are measured at all times. Additionally, the scanning mechanism in an FTS dictates the maximum spectral resolution that can be achieved in an FTS. FTS scanning mechanisms with sufficient performance for $R\geq1000$ have substantial heritage from terrestrial, balloon, and space based platforms~\cite{Gom2014,Griffin2010,Carli84}.

On-chip filterbank spectrometers (FBSs) are a burgeoning technology that provides compact spectral dispersion at the focal plane through superconducting resonators~\cite{Kovacs2012,Endo2012,Barry2022}. Each FBS device couples light from an ``on-sky'' antenna to a series of spectral filters that each terminate in a microwave kinetic inductance detector (MKID). In principle, several FBSs can be fielded to a focal-plane array in which each on-sky pixel is connected to a spectrometer. Increasing the resolution of an FBS necessitates an increase in the number of MKIDs, scaling as $\sim2R$ per polarisation (depending on spectral bandwidth and total designed spectrometer efficiency)~\cite{Barry2014PHD}. Thus, a focal plane of dual-polarisation $R\sim1000$ FBSs requires an additional factor of $\sim$3000--4000 detectors over a photometric focal plane. 

\begin{figure*}[!t]
    \centering
    \includegraphics[width=\linewidth]{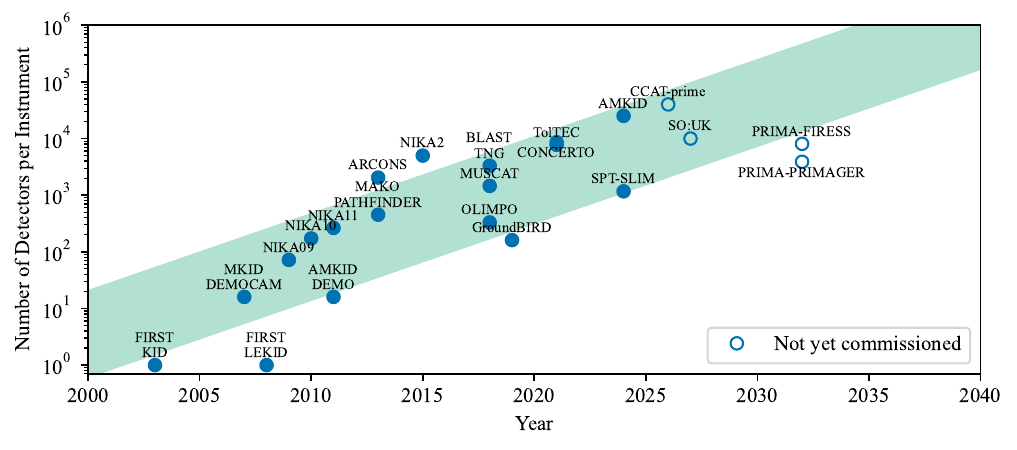}
    \caption{The total number of MKIDs employed at the focal plane of various instruments for astronomy. The shaded green area shows a linear fit to the data with a preferential weighting scheme in which more recent/future focal planes are assigned weights that are as much as twice that of early focal planes.}
    \label{fig:DetCount}
\end{figure*}

The number of MKIDs that can be fielded to a focal plane requires navigating a balance between the number of tones that can be read out on a single line and the thermal loading caused by these lines to the focal plane and their signal amplification~\cite{Sinclair2022}. The readout performance itself is largely dictated by the total power that can be divided and delivered to each tone while retaining sufficient signal-to-noise when read out by an analogue-to-digital converter~\cite{McHugh2012}. Significant advances in MKID detector counts are often a result of generational improvements in digital signal processing hardware and/or increases in detector responsivity that alleviate the level of amplification necessary. This provides an additional challenge for FBSs as increasing the resolution of a spectrometer results in a decreased spectral bandwidth, thereby increasing the detector sensitivity requirements to be photon-noise-limited. Fig.~\ref{fig:DetCount} demonstrates a rough trend in the progress of detector counts fielded to astronomical instruments which suggests that the technological developments needed to achieve the detector counts required for even modest arrays of medium-resolution FBSs may require $\sim$10--25 years.

\begin{figure}
    \centering
    \includegraphics[width=\linewidth]{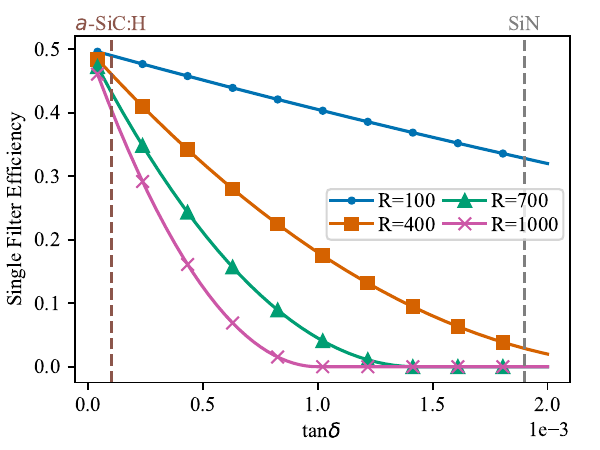}
    \caption{Efficiency loss in a single half-wave filter as a consequence of dielectric loss. Decreasing the bandpass of the filter (increasing its quality factor/$R$) results in greater efficiency loss.}
    \label{fig:DielectricLoss}
\end{figure}

FBSs also experience a loss in the total efficiency of a filterbank spectrometer due to the dielectric loss tangent of the materials used in the superconducting circuitry. This loss is augmented at higher spectrometer resolutions. Fig.~\ref{fig:DielectricLoss} shows that the efficiency of a single half-wave filter drops significantly with tan$\delta$ and as the designed resolution increases. For an FBS employing similar SiN to what was used in the South Pole Telescope Shirokoff Line Intensity Mapper (SPT-SLIM), sufficiently low loss to allow scaling to $R\sim1000$ has not been demonstrated ~\cite{Pan2023,Gethin2024PHD}. Low-loss dielectrics, such as $a$-SiC:H, have been demonstrated to have sufficiently low loss in the $\sim$1\,mm regime to significantly reduce this efficiency issue but have yet to be incorporated into multipixel FBSs~\cite{Buijtendorp2022}.

The fabrication of FBSs also becomes a significant barrier to reaching $R\sim1000$. In a typical lumped-element arrangement, the central frequency of FBS spectral channels is largely dictated by the length of the half-wave filter~\cite{Gethin2024PHD}. Thus, this imposes a limit on fabrication tolerances to develop a spectrometer that can effectively and efficiently sample a given spectral bandpass. For mm wavelengths, an R$\sim$1000 FBS would require feature variation that is less than $\sim$50\,nm~\cite{Thoen2022}. While this level of accuracy has been successful in the fabrication of a single FBS, scaling these techniques across an array of several hundred to thousand FBSs has yet to be demonstrated.

As a consequence of all of these challenges, FBSs that are currently being deployed have spectral resolving powers ranging from $R\sim100-700$ and only a few spatial/spectral pixels (1-15). To achieve an effective mapping spectrometer capable of $R\sim1000$ in the near term, we propose coupling a high-$R$ FTS with an FBS serving as a post-dispersing mechanism.

\section{Post-Dispersed FTS}
\begin{figure}
    \centering
    \includegraphics[width=\linewidth]{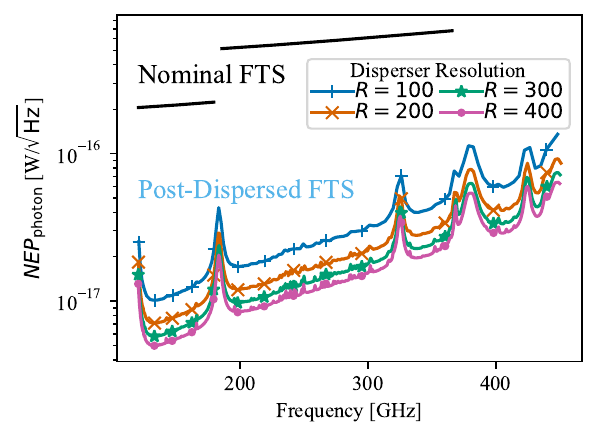}
    \caption{The reduction in the photon noise of an FTS provided by a low-resolution post-dispersion spectrometer. The optical load estimate has been determined from a model of a 15m ground-based observatory at Maunakea. A 70\% efficiency for the post dispersing spectrometer has been assumed.}
    \label{fig:NEPreduce}
\end{figure}
A post-dispersed FTS combines a traditional medium-$R$ FTS with a low-$R$ dispersing spectrometer to reduce the band incident on each detector. In this arrangement, the resolution of the spectrometer is entirely determined by the FTS, while a dispersing spectrometer with an $R$ of a few hundred is capable of reducing the photon noise incident on the instrument's detectors by over an order of magnitude. This is demonstrated in Fig.~\ref{fig:NEPreduce} comparing against two representative FTS bands. Here we note that a single FBS focal plane employing half-wave resonators is capable a full octave of bandwidth with band centres chosen to reduce load from atmospheric features~\cite{Barry2022}. The optical load used in this Fig.~\ref{fig:NEPreduce} determined using a model of a 15-m ground-based telescope coupled to an FTS with warm optics under a 25th percentile weather condition at Maunakea. A 70\% efficient backend dispersing spectrometer has been chosen as it is representative of the maximum theoretical efficiency of half-wave FBS and of similar efficiency to current generation sub-mm grating spectrometers~\cite{vieira2019,Barry2014PHD}. In addition to this reduction in noise, displacing the burden of medium resolution spectroscopy to an FTS reduces the detector count required for such an imaging spectrometer by an order of magnitude over that of only an FBS array.

\begin{figure}
\centering
    \includegraphics[width=0.45\linewidth]{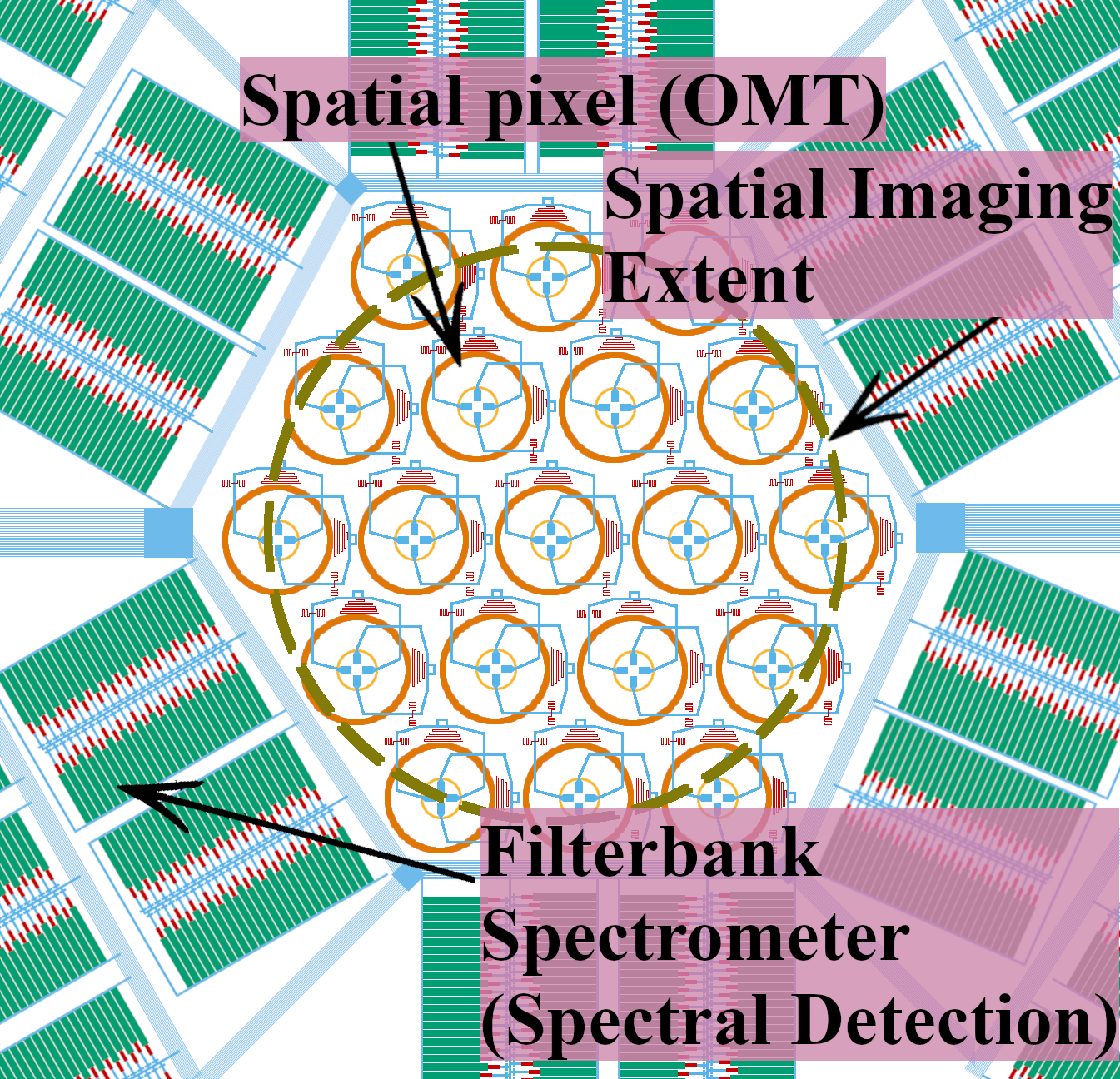}
    \includegraphics[width=0.45\linewidth]{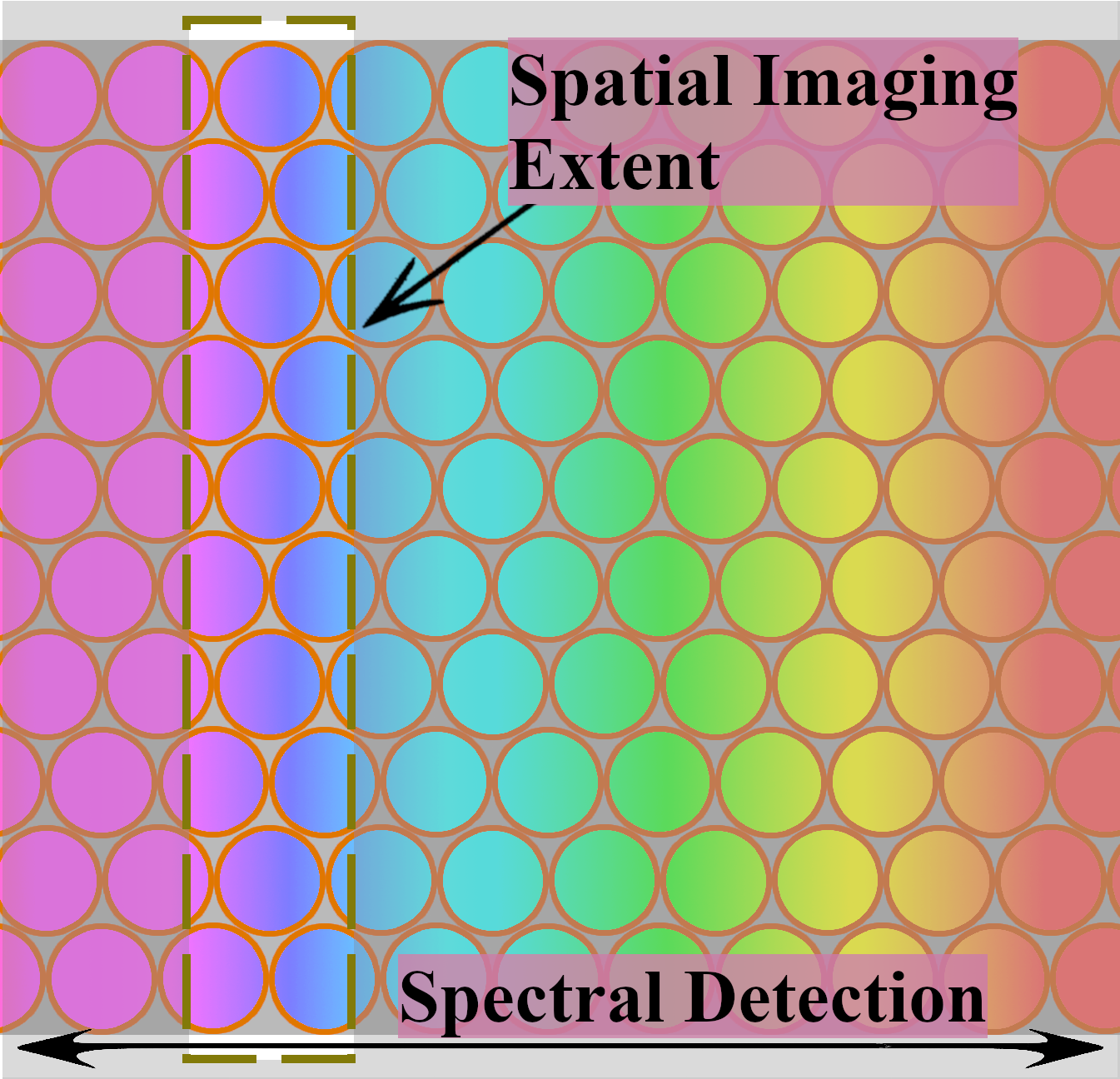}
    \caption{The focal plane of an FBDFTS (left) and that of a grating-dispersed FTS (right, adapted from Bradford et al.~\cite{Bradford2025}) are shown diagrammatically.}
    \label{fig:focalPlanes}
\end{figure}

Post-dispersed FTSs employing a low-R grating spectrometer have seen some interest, leading to instrument designs and laboratory-based demonstration through the SPICA-SAFARI and PRIMA-FIRESS mission proposals/studies~\cite{Naylor2022,Bradford2024,Bradford2025}. In particular, Bradford et al. (2025)\cite{Bradford2025} note the advantages a post-dispersed FTS instrument has over other spectrometer designs with similar performance targets (including FPS-based designs). The use of an FBS as a post-dispersion spectrometer provides particular advantages over more traditional spatially dispersing architectures:
\begin{itemize}
\item With an FBS, the spectral dispersion occurs over a total footprint that is about a centimetre in length at $R\sim200$ while the illumination area of a grating spectrometer scales significantly with wavelength. As such, mm/sub-mm gratings of similar resolution are of the order $\sim$1\,m in their size. Additionally, coupling an FTS to a grating requires additional optics (including an intermediate focus), significantly increasing the size and optical complexity of the post-dispersed FTS~\cite{Pastor2016}.
\item Light's spectral content is not spatially dispersed prior to the focal plane. Therefore, there is no significant compromise to the fundamental imaging capabilities of the FTS. A grating disperser suffers from the same imaging disadvantages outlined in the previous section. Again, we note that a grating spectrometer sacrifices a spatial-imaging dimension in order to measure spectral content. In principle, an FPS can displace spectral dispersion away from a 2-dimensional imaging array of antennae (see Fig.~\ref{fig:focalPlanes} and Barry et al 2022~\cite{Barry2022}).
\item Traditional dispersive elements often have polarisation dependencies. As a consequence, pairs of dispersion elements may be needed to recover the full efficiency of the FTS. With an FBS disperser, polarisation can be handled at the focal plane using common polarimeter designs and techniques. E.g., coupling a dual-polarisation orthomode transducer antenna to two filterbanks, one for each polarisation~\cite{Barry2022,Yoon2009}.
\end{itemize}

\section{Filterbank-dispersed FTS}
\label{sec:FBDFTS}
\begin{figure*}[!t]
    \centering
    \includegraphics[width=0.9\linewidth]{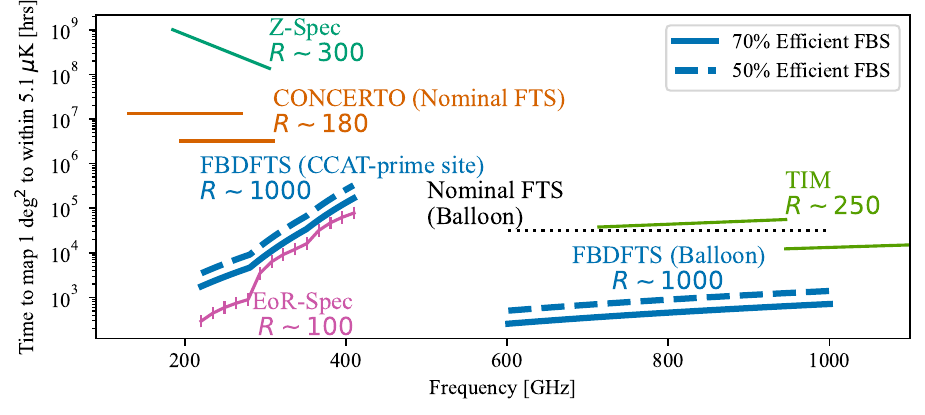}
    \caption{The projected mapping speed of an FBDFTS from the CCAT-prime/FYST site and from a balloon platform compared to other prominent mapping spectrometers in similar spectral bands. The instantaneous bandwidth of heterodyne spectrometers is denoted by vertical bars subdividing their respective curves. Mapping speeds have been determined using published sensitivity values and are discussed more fully in Section~\ref{sec:FBDFTS}.}
    \label{fig:MappingSpeeds}
\end{figure*}
Currently, there is significant potential for a filterbank-dispersed FTS (FBDFTS) instrument as both $R\sim1000$ mm/sub-mm FTSs and multipixel FBSs with an $R\sim100$ have both been demonstrated on-sky through SPT-SLIM and FTS-2, respectively~\cite{Karia2025,Gom2010}. FTSs have significant heritage of providing large-FoV imaging spectroscopy for mm/sub-mm astronomy and we note that the $\sim$2\,arcmin FoV FTS-2 and $\sim$20\,arcmin CarbON CII line in post-rEionisation and ReionisaTiOn epoch both suffered some degree of FoV compromise due to the size constraints from FTS-2's location in the JCMT optical path and CONCERTO's ambitious scanning speed~\cite{Gom2007, 2020Concerto}. Additionally, the number of detectors employed by a fully band-covered FBS is given by,
\begin{equation}
    N_\mathrm{det} = \Sigma R \log\left(\frac{\nu_\mathrm{max}}{\nu_\mathrm{min}}\right),
    \label{eq:numDets}
\end{equation}
where $\log$ is the natural logarithm, $\Sigma$ is an oversampling factor to increase the total efficiency of the spectrometer (normally $\Sigma = 1.5$--2), and the spectral band of the spectrometer is defined by the frequencies $\nu_\mathrm{max}$ and $\nu_\mathrm{min}$. As such, an FBDFTS with $10^2$--$10^3$ spatial/spectral pixels and an $R\sim200$ FBS disperser is well within the expected detector counts of near-term mm/sub-mm instruments. 

To demonstrate the imaging capabilities of an FBDFTS, we have calculated the mapping speed of an FBDFTS instrument from both a ground-based platform (CCAT-prime/FYST observatory) and from a balloon-borne telescope similar to the proposed BLAST observatory (see Fig.~\ref{fig:MappingSpeeds})~\cite{2023CCAT,2024BLAST}. Other mapping spectrometers have also been included for comparison using reported values of noise equivalent flux densities and optical loads from their respective exposure time calculators and/or from literature (zenith pointing under ideal weather conditions)~\cite{zspecNEFD,2020Concerto,TIMMapping,2023CCAT}. We note that for EoR-Spec, its instantaneous bandwidth is marked by vertical bars and the time penalty required to measure its full band has been included in the calculation. For the FBDFTSs, we have considered a 50\% efficient FTS coupled to an $R=200$ FBS with half-wave filters. Though a single half-wave filter can be at most 50\% efficient, FBSs can employ a spectral oversampling technique to improve their theoretical efficiencies to $\sim$70\%~\cite{Barry2014PHD,Gethin2024PHD}. From both platforms, an FBDFTS has the potential to provide an orders-of-magnitude improvement over a nominal FTS and an extremely competitive medium-resolution mapping spectrometer, reaching similar mapping speeds to low-resolution designs. For a ground-based FBDFTS, the CONCERTO FTS serves as a natural point of comparison, while a nominal FTS comparison has been determined for the balloon-borne FBDFTS performance. Of particular note, an FBDFTS has the potential of providing similar mapping speeds to EoR-Spec with an order of magnitude improvement in spectral resolution. To achieve a similar level of resolution with the EoR-Spec FPS, the instantaneous bandwidth would also be reduced by the same factor and result in an increased time penalty to fully scan the spectral band.

We note some additional subtleties that can arise from separating the dual output ports of an FTS as in the Mach-Zehnder design~\cite{Naylor2003}. Since such an FTS naturally lends itself well to two focal planes, dramatic improvements over the projected performance in Fig.~\ref{fig:MappingSpeeds} and/or practical improvements beyond what we have discussed can be obtained. (1) Employing a duplicate FBS focal plane at the second FTS output, would make it possible to recover a near-100\% efficient FTS. (2) A second FBS focal plane designed around a different band could be used to increase the spectral bandwidth of the instrument to two octaves~\cite{Griffin2010}. (3) The second import can be place off-axis on the sky to provide natural background subtraction~\cite{Gom2007}. (4) Polarisation could be separated in the FTS's interferometer portion, terminating on separate focal planes dedicated to orthogonal polarisation states~\cite{Nagler2015}. In many of these and similar designs, one could consider a single focal plane  instrument as baseline/demonstrator that could later be upgraded.

\begin{figure}
    \centering
    \includegraphics[width=\linewidth]{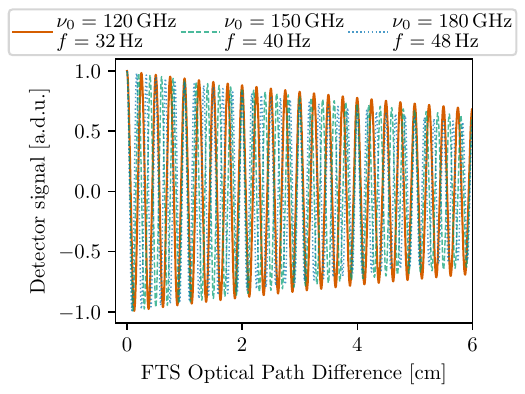}
    \caption{Simulated detector signals from continuum measurements with three spectral channels of an FBDFTS. The low-frequency channel (with a spectral band centred at $\nu=120$\,GHz, marked by the heavily weighted orange line) emphasises the decaying cosinusoidal modulation that would be measured by a single detector in an FBDFTS. The signal frequency, $f$, in the temporal domain is set by the scanning speed of the FTS mirror ($2\,$cm/s) in the FBDFTS and the central frequency of the filter channel.}
    \label{fig:examplesigs}
\end{figure}

An FBDFTS also provides notable practical benefits that are capable of further reducing noise in the signal domain. An FBDFTS exploits the nominal potential of an FTS to effectively select the modulation frequency of the measured interference signal, being dependent on the spectral band measured and the speed at which the interferometer is scanned. This allows for a practical instrument solution to provide signal modulation in preferential signal bands, e.g., avoiding $1/f$-noise dominated regions. Coupled with this, the narrow-banded profile of FBS filters maps to a narrow-banded signal measured in the time domain by the detector. As such, the measurement signal from an FBDFTS instrument is well-positioned for signal filtering, amplification, and advanced processing techniques. As an example, Fig.~\ref{fig:examplesigs} shows the theoretical interference signal from an FBDFTS measured in three of its filters (each with $R=200$) while viewing a continuum spectral profile in their respective bands~\cite{Barry2022}. Fig.~\ref{fig:examplesigs} also notes the signal modulation frequency for each example FBS channel in a Mach-Zehnder-style FTS with a scanning speed of $2\,$cm/s, being $\sim$30--50\,Hz~\cite{2003Naylor}. Mechanical speeds of this order are easily achievable with modern translation stage hardware and the scanning speed becomes a tradeoff between atmospheric effects and vibrational noise introduced to the optical system~\cite{2020Concerto,2022Concerto}.

\section{Potential for an FBDFTS LIM experiment}
\label{sec:LIMExp}
\begin{figure*}
    \centering
    \includegraphics[width=\linewidth]{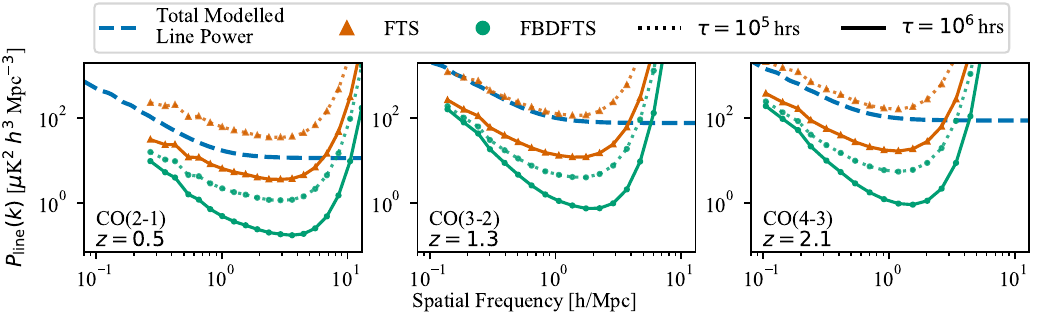}
    \caption{The predicted performance of an $R=1000$ LIM experiment leveraging a photon-noise limited FBDFTS operating at 150\,GHz from the JCMT is compared against the performance of a typical FTS for the rotational transitions: CO(2-1) at $z=0.5$ (left), CO(3-2) at $z=1.3$ (centre), and CO(4-3) at $z=2.1$ (right). Integration times of $10^5$ and $10^6$ spectrometer hours are both shown as conservative/ideal-case scenarios for the experiment. The modeled power spectra shown are taken from Karkare et al. (2022)~\cite{Karkare2022}. To determine loading from the sky, a model of the Maunakea atmosphere above the JCMT while zenith pointing was used under its 50th percentile weather conditions (1.86\,mm precipitable water vapour). Additional details on the atmospheric model and the optical model used to determine optical loading are given in Section~\ref{sec:LIMExp}.}
    \label{fig:LIMproj}
\end{figure*}

An FBDFTS has substantial potential to provide, to our knowledge, the first $R\sim1000$ sub-mm/mm LIM experiment in the near-term. Scaling LIM measurements to higher $R$ allows for the recovery of higher spatial modes in the LIM power spectrum and allows for the spatial resolution of galaxy sources parallel to the line of sight. Additionally, improved spectral/redshift resolution assists in the extraction of contaminant spectral features~\cite{2015Sliva,2016Lidz}.

Since the sensitivity of LIM experiments is a consequence of the number of on-sky spatial/spectral pixels, their effort level/integration time is commonly expressed as ``spectrometer hours'' ($\tau$); i.e., the product of integration time and the number of spatial/spectral pixels. Scaling FBS focal planes from their current sizes of a few 10s of pixels to $\sim$100 is expected within the next few years~\cite{2024Rybak}. As such, we expect a LIM experiment leveraging an FBDFTS to be well capable of $10^5$--$10^6$ spectrometer-hour surveys. 

Due to the rigour and availability of SCUBA-2's medium-R FTS-2 optical model and optical loading estimates~\cite{Gom2010,Naylor2003}, we have considered a photon-noise limited FBDFTS deployed from the James Clerk Maxwell Telescope (JCMT) employing a similar optical configuration to predict the LIM performance of such an instrument. We have considered a room-temperature Mach-Zehnder $R=1000$ FTS operating at 150\,GHz employing an $R=200$ FBS dispersing focal plane with 100 spatial/spectral pixels. An instrument of this type would require $\sim$$4.5\times10^{4}$ detectors in total and $\sim$450\,mm of mechanical travel for the moving mirror in the interferometer. We have taken the RMS surface roughness of the primary mirror and secondary mirror to be 20\,$\mu$m and 15\,$\mu$m, respectively, and all relay optics and FTS optics to have a roughness of 1\,$\mu$m. The atmospheric loading has been determined using a model from the atmospheric modelling software, \textit{am}, for Mauna Kea in its 50th percentile water vapour conditions (1.86\,mm precipitable water vapour), at zenith pointing~\cite{PaineAM}. Following the methodology of the CONCERTO collaboration~\cite{2020Concerto}, we have determined the expected constraints provided by an FBDFTS LIM experiment on the CO(2-1), CO(2-3), and CO(4-3) power spectra. These are shown in Fig~\ref{fig:LIMproj}. In each case, we have also shown predicted LIM power spectra from clustering and shot powers from Karkare et al. (2022)~\cite{Karkare2022}. This demonstrates that the proposed FBDFTS is capable of an orders-of-magnitude improvement over a nominal FTS in a LIM experiment and is capable of LIM measurements of signal to noise 10-100 within near-term survey times ($\sim$1000 hrs).

Recently, the CONCERTO LIM experiment represented a significant effort to measure the power spectrum from ionised carbon through a 840\,h survey at the APEX telescope with a 1720 pixel FTS (i.e., $\sim$10$^{6}$ spectrometer hours). Based upon anticipated optical loads and laboratory characterisation of the detector array, CONCERTO projected an LIM measurement of CII with a signal-to-noise ratio of 23--2 at $z=5.5$--$8$ with $R\sim300$. As the noise in the LIM signal approximately scales with the ratio of the detector's spectral band ($\Delta \nu$) and the frequency bin-width of the spectrometer's resolution element ($\delta \nu$), we project a factor of $\sim$15 improvement over the maximum theoretical LIM performance of CONCERTO with an $R=1000$ FBDFTS employing an $R=200$ dispersing filterbank (with the approximation $NEP_\mathrm{photon} \propto\sqrt{\Delta\nu})$.

Despite this promising performance, the $R\sim1000$ FTS portion of the proposed FBDFTS does introduce a significant source of optical complexity and systematics.  Though challenging, there is substantial heritage in the characterisation and mitigation of systematics in FTS instruments employed for astronomical measurements through both ground-based and on-sky measurements~\cite{2010Locke,2014Swinyard,2015Hopwood,2018Valtchanov,Gom2010,2013Makiwa,Gom2014}. Additionally, the FTS can in many respects be considered a self-calibrator for the FBS portion of the instrument, as FTSs have significant heritage in providing spectral response measurements with excellent frequency calibration~\cite{2004Locke,2010Pajot,2014Hailey,2014Karkare}. The commissioning of an FBDFTS LIM experiment would necessarily require a considerable calibration program.
\section{Conclusion}
Current sub-mm/mm mapping medium-R spectrometer architectures all suffer from significant drawbacks that limit their scalability to the current generation of large-FoV spectral galaxy surveys and LIM. Using a low-resolution spectrometer as a post-dispersing element in a higher resolution FTS is capable of reducing the relatively high photon noise of FTSs by more than an order of magnitude. Employing an FBS dispersing focal plane retains the mapping advantages that have historically been sacrificed by more traditional dispersing spectrometers in post-dispersed FTS instruments.

We have shown that this noise reduction can result in an orders-of-magnitude improvement in the mapping capabilities over a nominal FTS. We have also demonstrated the considerable potential to achieve an $R\sim1000$ FBDFTS mm/sub-mm LIM experiment in the near-term. We have demonstrated that an FBDFTS deployed from the JCMT would be capable of measuring the power spectra of CO(2-1), CO(3-2), and CO(4-3) features at $z=0.5$, $z=1.3$, and $z=2.1$, respectively, with a signal to noise ratio of $\sim$10--100 within $10^5$--$10^6$ spectrometer hours.

\section{Acknowledgments}
This work is supported by UKRI Future Leaders Fellowship MR/W006499/1.

\bibliography{refs} 
\bibliographystyle{IEEEtran}

\newpage

\section{Biography Section}
 
\begin{IEEEbiography}
[{\includegraphics[width=1in,height=1.25in,clip,keepaspectratio]{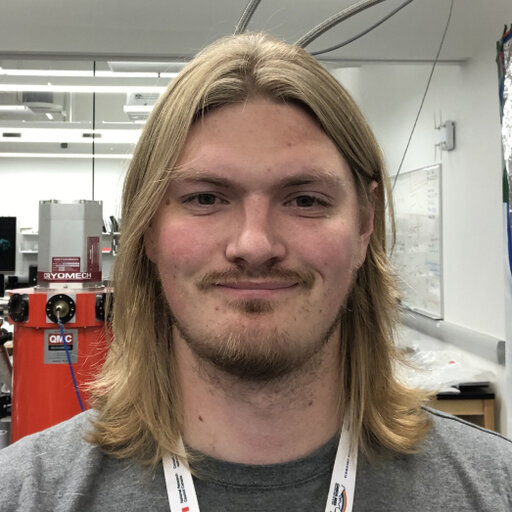}}]{Chris S. Benson.}
is a research associate at Cardiff University. He received his B.S. (hons.), M.Sc., and Ph.D. degrees in physics and experimental astrophysics from the University of Lethbridge, Canada in 2018, 2020, and 2024, respectively. He has worked on Herschel SPIRE calibration, the development of a Double-Fourier interferometry testbed and the characterisation of its associated detector array, and is working on the spectrometer focal plane of the South Pole Telescope Shirokoff Line Intensity Mapper. His current research interests include on-chip spectrometers, Fourier transform spectroscopy, double-Fourier interferometry, and mm-wave and terahertz detectors.

\end{IEEEbiography}

\vspace{11pt}


\vfill

\end{document}